\documentclass[fleqn,10pt]{wlscirep}
\usepackage[utf8]{inputenc}
\usepackage[T1]{fontenc}
\usepackage{makecell}
\usepackage{multirow}

\title{A large synthetic dataset for machine learning applications in power transmission grids}

\author[1,*]{Marc Gillioz}
\author[1]{Guillaume Dubuis}
\author[1,2]{Philippe Jacquod}
\affil[1]{School of Engineering, University of Applied Sciences and Arts Western Switzerland HES-SO, 1951 Sion, Switzerland}
\affil[2]{Department of Quantum Matter Physics, University of Geneva, CH-1211 Geneva, Switzerland}
\affil[*]{corresponding author: marc.gillioz@hevs.ch}

\DeclareMathOperator{\var}{var}

\begin{abstract}
With the ongoing energy transition, power grids are evolving fast. They operate more and more often close to their technical limit, under more and more volatile conditions. Fast, essentially real-time computational approaches to evaluate their operational safety, stability and reliability are therefore highly desirable. Machine Learning methods have been advocated to solve this challenge, however they are heavy consumers of training and testing data, while historical operational data for real-world power grids are hard if not impossible to access. This manuscript presents a large synthetic dataset of power injections in an electric transmission grid model of continental Europe, and describes the algorithm developed for its generation. The method allows one to generate arbitrarily large time series from the knowledge of the grid -- the admittance of its lines as well as the location, type and capacity of its power generators -- and aggregated power consumption data, such as the national load data given by ENTSO-E. The obtained datasets are statistically validated against real-world data. 
\end{abstract}
\begin{document}

\flushbottom
\maketitle

\thispagestyle{empty}

\begin{quote}
\emph{
This version of the article has been accepted for publication, after peer review but is not the Version of Record and does not reflect post-acceptance improvements, or any corrections. The Version of Record is available online at: \href{https://doi.org/10.1038/s41597-025-04479-x}{https://doi.org/10.1038/s41597-025-04479-x}}
\end{quote}

\section*{Background \& Summary}

Electric power grids are fast-evolving, with the massive integration of new renewable energy resources -- either distributed over lower voltage levels, or as large offshore wind farms, multi-GW solar photovoltaic parks or GW battery energy storage parks -- and with the accompanying necessary grid upgrades. Incorporating these new resources into the legacy grid poses significant challenges to grid operators. Power grids operate more often closer to their technical limit, while undergoing larger, more frequent fluctuations. Evaluating the safety and reliability of operation needs to be done more often -- optimally in real-time -- and with better accuracy. This is becoming computationally challenging. Fast, data-based methods are often advocated as solutions to this problem. In particular, Machine Learning (ML) approaches are of interest, because they learn efficiently and provide results on fast timescales.\cite{donnot2017, arteaga2019, fioretto2020, duchesne2020, misyris2020} ML algorithms are heavy consumers of training and testing data, however, and the need for large, high-quality sets of data has been identified as one of the key challenges to overcome, when constructing ML algorithm ready for real-world application.\cite{Stiasny2022}

There is thus a clear need for open-access, large sets of operational data of electric power grids for ML applications. 
Historical operational data are however notoriously hard, if not impossible, to obtain.
For this reason, most studies rely on synthetic datasets that span the entire feasible space of grid configurations and focus on representative operational conditions by appropriate sampling.\cite{krishnan2011, hamon2016, konstantelos2018, thams2019, venzke2021, joswig2022, rossi2022data, bugaje2023generating, perbellini2024}
In all such cases, the state of the grid is realistic at any given data point, but the historical structure and correlations between power injections at different times are lost.
On the other hand, time series analyses have been proven powerful, but they are generally limited to small systems because of the lack of good-quality data.\cite{Zhang2021}
In fact, real-life time series exist in aggregated form, e.g. with time-resolved power generation and consumption at national level.
However, geographical resolution at the level of individual buses is highly desirable and actually necessary for detailed power flow analysis. Existing databases with such resolution are based on synthetic data.\cite{simbench,pypsa1,pypsa2} They are helpful for specific power flow investigations such as operational studies, generation and transmission expansion studies or grid extension planning. They contain however too many correlations between time series of power injections at different buses to be of any help for ML applications. As a matter of fact, training and testing a ML algorithm on artificially correlated ground truth data inevitably leads to biases that will hinder the successful deployment of the algorithm in real-world situations.

Large, realistic, high-quality sets of ground truth data are therefore needed for ML applications. By realistic, we take the definition that, for a given model of a real-world power grid, the statistical properties of the ground truth dataset -- in particular the average and variance of the injections at each bus, the typical daily, weekly and seasonal periodicities, the cross-correlations between power injections at two different buses and so forth -- need to reflect those
of available sets of real-world data. 
The purpose of this paper is to meet the need of ground truth datasets for ML applications. We construct and describe a generic algorithm to generate realistic datasets for different grids of varying sizes, with possibly different power generation mixes and consumption profiles. Our algorithm generates arbitrarily large time series of power injections -- generation and consumption -- from time-resolved, aggregated data. In the
first step, power consumption data are disaggregated over all consumer buses, following a procedure that takes into account the local demography in the vicinity of each bus, while introducing a tunable noise contribution that reduces the cross-correlation between different buses to the desired level. To still capture the daily, weekly and seasonal periodicities of power consumption, the procedure incorporates noise in the Fourier components of the demographically disaggregated power time series -- not on the time series themselves. The resulting consumption is then fed to an Optimal Power Flow (OPF) that delivers generation time series. To avoid systematic line loads at 100 \%, the 
thermal limits of the power lines are incorporated in the objective function of the OPF, and not as a rigid constraint.

The algorithm is applied to the PanTaGruEl model of the synchronous electric power grid of continental Europe,\cite{PanTaGruEl} with data from the publicly available ENTSO-E transparency platform
(\href{https://transparency.entsoe.eu/}{https://transparency.entsoe.eu/}). The platform is managed by the European Network of Transmission System Operators for Electricity, which collects  data directly coming from the operators of its member countries, including in particular countrywise-aggregated time series of consumption data.
The output of the algorithm is a set of 20 years of sequential data with a time resolution of one hour for several thousands of loads and several hundreds of generators of various types.
To the best of our knowledge, this is the first dataset of this type. It will serve as a basis for time series anomaly detection in transmission grids in a future work of ours.

It is worth noting that the algorithm has sufficient flexibility that it allows one to tune datasets closer to the $N$ or $N-1$ feasibility boundary, e.g. for
more efficient training, depending on the desired ML application. The method is exportable to any other grid model, provided aggregated power injection data are available. In particular, it can easily account for national or regional subsets of PanTaGruEl.

\section*{Methods}

The methods have been developed to obtain a realistic model of power transmission grids, generating datasets for power injection that are both
\begin{itemize}
    \item coherent in space, meaning that the power consumed in one place is mostly produced by nearby generators,
    \item coherent in time, meaning that power generations and consumptions have average values and fluctuations that are similar to real-world data, including their daily, weekly and seasonal periodicities, as well as their correlations.
\end{itemize}
We start with a model of the transmission grid of continental Europe, and open-access aggregated real-world time series for production and consumption. Next, we generate disaggregated, synthetic time series for consumption on each bus, which we distribute  on the grid. Finally, we design and solve an optimization problem that, for each time step, creates a realistic dispatch for power generators.

\subsection*{Model of the European transmission grid}

\begin{figure}[ht]
    \centering
    \includegraphics[width=0.8\linewidth]{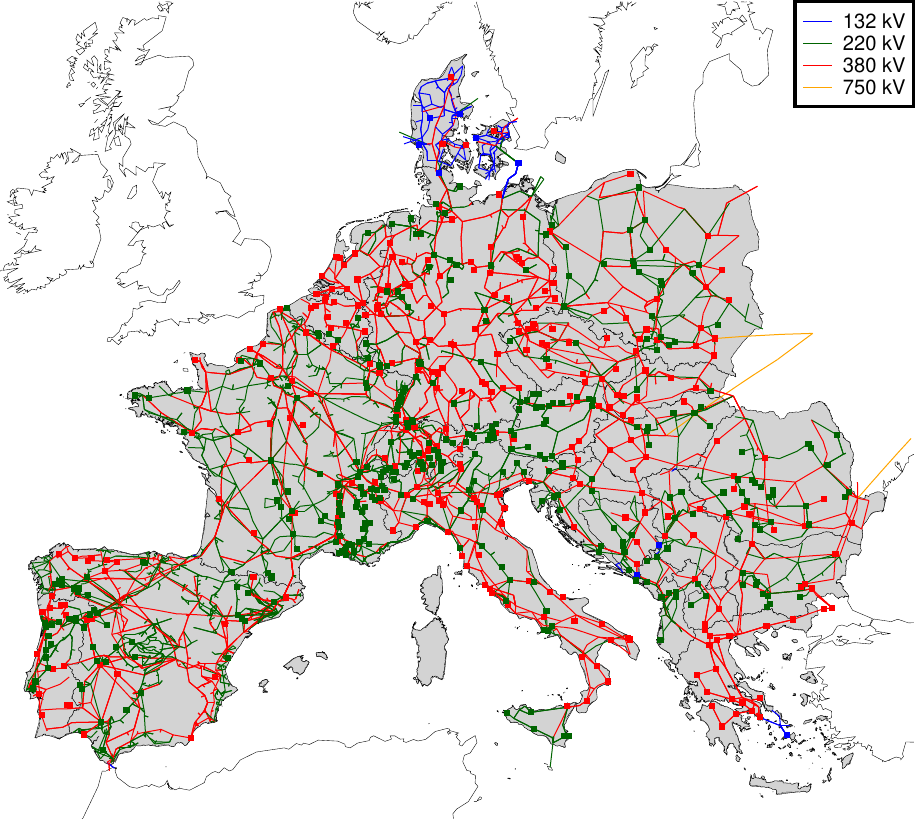}
    \caption{The PanTaGruEl model of the transmission grid of continental Europe.\cite{PanTaGruEl}
    Lines are colored by voltage. Buses with generators attached are indicated with a square.}
    \label{fig:network}
\end{figure}

For the transmission grid, we use the PanTaGruEl model of the synchronous electric grid of continental Europe\cite{PanTaGruEl} in its most recent version as a Julia package (\href{https://github.com/laurentpagnier/PanTaGruEl.jl/}{https://github.com/laurentpagnier/PanTaGruEl.jl/}) exported in PowerModels format.\cite{PowerModels}
The model consists of 7822 power lines and 553 transformers connecting 4097 buses, to which are attached 1083 generators. It is illustrated in Fig.~\ref{fig:network}.
The model includes power line admittances, as well as location, type and maximum capacity for all generators. Additionally, it attributes a weight factor to each bus with a load, giving the ratio of the national load consumed by that bus. Weight factors are estimated based on the local demographics in the vicinity of each bus. 
Details of the model and how it has been constructed can be 
found in Refs.~\cite{PagnierJacquod2019, TylooPagnierJacquod2019}

The power grid model gives a realistic representation of the synchronous grid of continental Europe. It 
has been validated in a number of studies. Ref.~\cite{PagnierJacquod2019} compares model-based and real-world 
installed generation capacities. Static properties such as load on power lines 
are qualitatively similar to real ones for various dispatches.\cite{TylooPagnierJacquod2019} Dynamic properties such
as disturbance propagation times,\cite{PagnierJacquod2019} frequency response amplitude to perturbations,\cite{Pagnier2019, Tyloo2021} as well as frequency and
geographical dependence of inter-area oscillations~\cite{Fritzsch2022}
have been found to qualitatively match those of the real synchronous grid of continental Europe. The model has been recently used to validate fast numerical algorithms for power flow and dynamic studies of electric power grids.~\cite{guichard1,guichard2}

Currently -- i.e. up to, but not including the work reported here -- the model is well-suited for investigations around a given operational state. It needs however to be corrected for ML and other applications requiring statistically realistic time series. In particular, 
two issues are fixed below, which have required special attention. First, the original dispatch algorithm disaggregates national loads on the consumption buses, according to the fixed, time-independent weight factors mentioned above. That procedure obviously generates unrealistic, perfectly correlated time series for consumption on 
all load buses in a given country, because they are all proportional to the national load. Second, the original OPF includes the thermal limit of each power line in a set of constraints on power flow solutions. Minimizing the cost function in the resulting non-convex optimization problem leads almost systematically to operational states lying on the boundary of the feasibility region, i.e. with one or several lines loaded at 100 \% of their thermal capacity. This also is not realistic and needs to be corrected 
for machine learning applications. 

\begin{figure}[ht]
    \centering
    \includegraphics[width=0.65\linewidth]{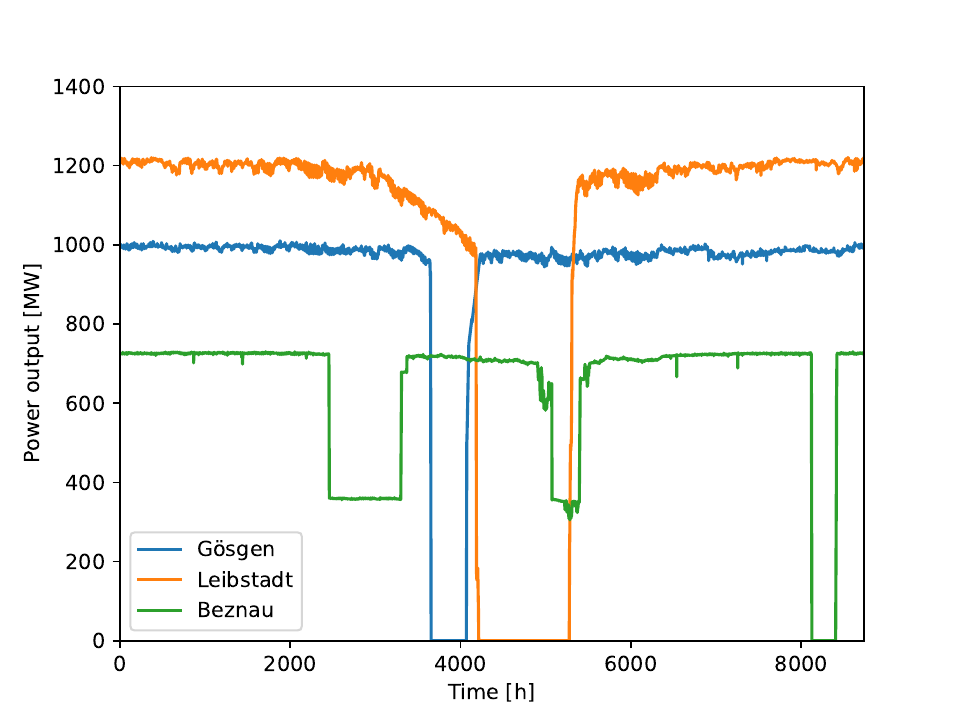}
    \caption{Historical time series for the three active Swiss nuclear power plants in 2020, obtained from the ENTSO-E transparency platform. 
    Nuclear generators mostly work at their nominal regime,
    except for periods of planned maintenance (mostly in summer, and alternating between plants)
    and rare unexpected events. The G\"osgen and Leibstadt power plants have a single reactor, while the Beznau power plant has two reactors,
    which explains why the power output is about half the nominal value during maintenance.}
    \label{fig:nuclear-series}
\end{figure}

Before we tackle these two main issues, minor, practical  
changes are made to PanTaGruEl.
First, we remove the 268 small generators whose capacity is below 50 MW, bringing the number of generators down to 815. The reasons for this are that (i) 
in most countries, generators with a rated power below 100 MW are connected to lower voltage levels, (ii) the removed generators account for a total capacity of 3.4 GW, i.e. about 0.7 \% of the total installed capacity of 466 GW in PanTaGruEl, and consequently (iii) dispatching such small generators does not affect the OPF solution significantly, but substantially increases computation times.
Second, generators are classified as follows:
\begin{itemize}
    \item The categorization of generators into various types in the ENTSO-E transparency platform differs from country to country. We reduce the number of different generator types in order to unify that categorization. The power generator types are given in Table~\ref{tab:model-gens}.
    
    \item We do not treat nuclear plants in the dispatch algorithm, instead they have power generation profiles matching historical, high-quality data available on the ENTSO-E transparency platform.
    This is appropriate, since nuclear reactors are only weakly flexible, with limited load-following capabilities. Accordingly, they operate mostly in an on/off mode, dictated by planned maintenance rather than by the electricity market or the operational state of the grid. Typical time series for power generation of nuclear plants are shown in Fig.~\ref{fig:nuclear-series}.
    
    \item An availability factor is attributed to all non-nuclear power plants, giving the percentage of time the plant would need to produce at its rated power to achieve its announced annual production.
\end{itemize}
All non-nuclear generators are treated as dispatchable in the OPF. 
Additionally, coal generators have a ramping constraint. Given that we use an 
hourly time resolution, there is no such constraint for hydro, nor gas-fired plants. 

Table~\ref{tab:model-gens} gives the number of generators, their rated capacity and their availability factors, listed by country and type.
Generation in the PanTaGruEl model is dominated by fossil fuels, followed by hydro. The few generators classified as \textit{other} are geothermal (1) and biomass (11) power plants, or of unknown type (16). Clearly, PanTaGruEl significantly underestimates the importance of new renewable energy sources, as it is based on 2016 data. We stress that our goal in this paper is not to present an up-to-date model of an existing transmission power grid, instead to construct a general algorithm for generation of realistic synthetic data for a given model of a power grid. We note in passing that increased penetration levels of new renewable energy sources can be incorporated by, first, adding such productions at geographically appropriate places and, second, treating them outside the dispatch algorithm, i.e. with power generation given, e.g. by historical meteorological data of wind and insolation. 

\subsection*{ENTSO-E data}

Our algorithm relies on the ENTSO-E transparency platform (\href{https://transparency.entsoe.eu/}{https://transparency.entsoe.eu/}) as its single source of input data. These are historical time series with hourly resolution. Specifically, we consider 
\begin{enumerate}
    \item[(1)] the national loads of the 25 countries covered by PanTaGruEl, from January 2015 to December 2023;
    
    \item[(2)] each nuclear power plant individually, from January 2016 to December 2020;
    
    \item[(3)] all other plants aggregated by type, from January to December 2016;

    \item[(4)] the physical power flows crossing all borders between countries, from January 2016 to December 2020.    
\end{enumerate}
While uniquely valuable, data provided by the ENTSO-E transparency 
platform suffer from several, occasionally significant inconsistencies.~\cite{Hirth2018} For instance, the generation output listed by unit  refers sometimes to individual generators, sometimes to a multi-machine power plant. The main difficulty has to do with the aggregation of production by type that is not fully consistent over the whole database, with different countries using different nomenclatures.
We address this problem by regrouping generators into the five types of Table~\ref{tab:model-gens}. Note that the {\it hydro} type is further subdivided into \textit{storage} and \textit{run-of-river} -- which have quite different operating modes -- while the \textit{unspecified} sub-type gathers hydroelectric power plants of unknown or mixed type. 
Another difficulty in the ENTSO-E datasets is missing data. 
For each of the four sets of input data mentioned above, the time interval considered corresponds to the longest available period with almost complete time series, and for which
missing data are distributed over sequences with only few consecutive time steps. They can therefore be reconstructed through linear interpolation. 

The set (1) of input data -- the national loads -- is disaggregated and distributed onto the load buses of 
the model, following a procedure discussed in the next chapter. The set (2) of 
data is used directly as input into the OPF as injection series for each
nuclear power plants. The set (3) of data is used to calculate an 
average annual energy production which needs to be achieved by each power plant. It 
will be incorporated into the OPF as a constraint. 
The set (4) of data is used to correct inconsistencies between power generation and consumption data. 
As a matter of fact, the total annual production resulting from the model's generators does not exactly balance the total load from the ENTSO-E data. We would like to re-normalize the loads so as to satisfy the sum rule 
\begin{equation}\label{eq:sumrule}
{\cal P}_\alpha - {\cal L}_\alpha - {\cal E}_\alpha = 0 \, , 
\end{equation}
relating power production ${\cal P}_\alpha > 0$, load ${\cal L}_\alpha > 0$ and 
export ${\cal E}_\alpha$, for each country labeled by $\alpha$, adopting the 
convention that ${\cal E}_\alpha>0$ indicates a net power export.
The national consumption time series of set (1) are therefore multiplied by an appropriate factor, listed
in Table~\ref{tab:border-flows}, together with the national, annually integrated import/export balances ${\cal E}_\alpha$.

The last data processing step is to shorten all annual time series to 364 days, i.e. exactly 52 weeks. Time series then become approximately periodic, with their last week being very similar to their first week. 
We make use of this periodicity to rotate the series and make them start on a Monday. In this way, all annual time series are uniformly formatted with a total length of 8736 hourly steps. Accordingly, they can be directly compared to one another. 

\subsection*{Synthetic time series for loads}

\begin{figure}[ht]
    \centering
    \includegraphics[width=0.6\linewidth]{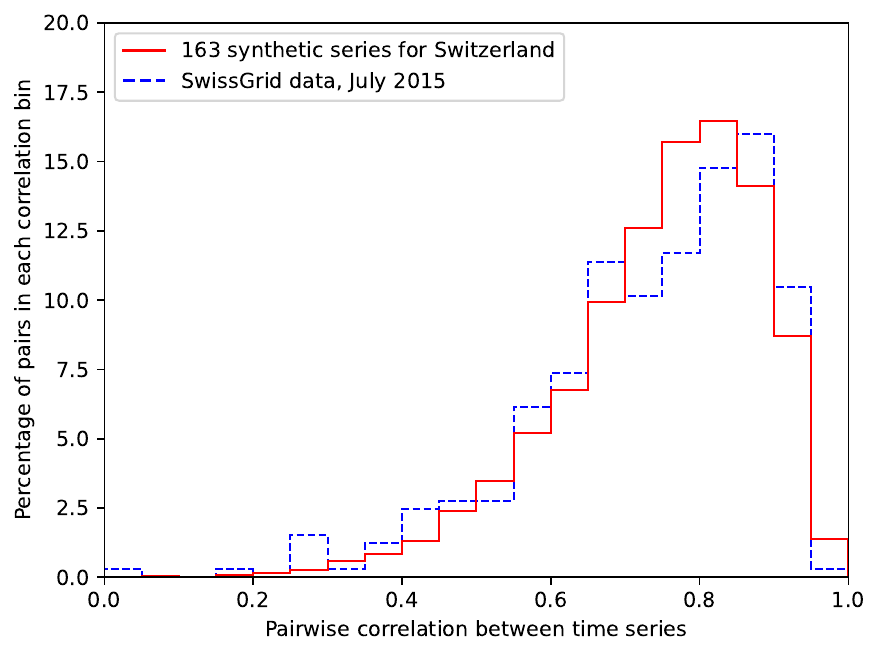}
    \caption{Distribution of Pearson correlation coefficients between pairs of
    synthetic time series, compared with real-world measurements for the Swiss transmission grid
    during the month of July 2015.}
    \label{fig:correlations}
\end{figure}

Our next task is to distribute the national load profiles obtained from the ENTSO-E transparency platform over the load buses, country by country. This is a highly non-trivial task, especially since we want to obtain realistic load profiles, particularly in terms of statistical distributions. One important ingredient to keep in mind, even more crucial for ML applications, is that time series corresponding to different load buses are often at least partially correlated. Our algorithm can tune the degree of correlation and a key validation of our method is that we are able to reproduce real-world distributions of Pearson correlation coefficients over the 163 nodes of the Swiss transmission grid. This is shown in Fig.~\ref{fig:correlations}.

The disaggregation method first builds statistical distributions of national loads for each country, using 9 years of ENTSO-E data from 2015 to 2023. Second, it 
draws time series of individual bus load from this distribution, weighted by the population density factor provided by PanTaGruEl. Using this factor is an approximation that  has been tested and validated, in that it leads to realistic flows.\cite{PagnierJacquod2019, TylooPagnierJacquod2019}

\begin{figure}[ht]
    \centering
    \includegraphics[width=0.7\linewidth]{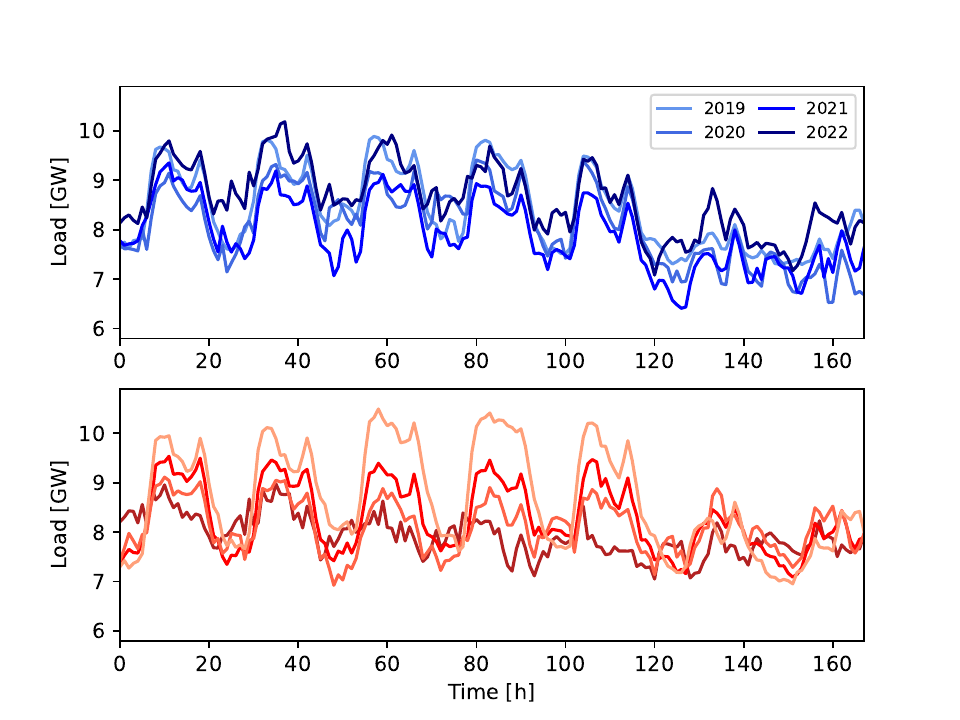}
    \caption{Total load for Switzerland during the fourth week of the year, with hourly resolution.
    The top panel shows historical data for the years 2019 to 2022.
    The lower panel shows synthetic series generated using a statistical model built from historical
    data. The same pattern is clearly visible in both panels:
    the five working days have higher load and two peaks in the morning and evening
    whereas the weekend is characterized by a lower consumption.}
    \label{fig:sample-time-series}
\end{figure}

\begin{figure}[ht]
    \centering
    \includegraphics[width=0.7\linewidth]{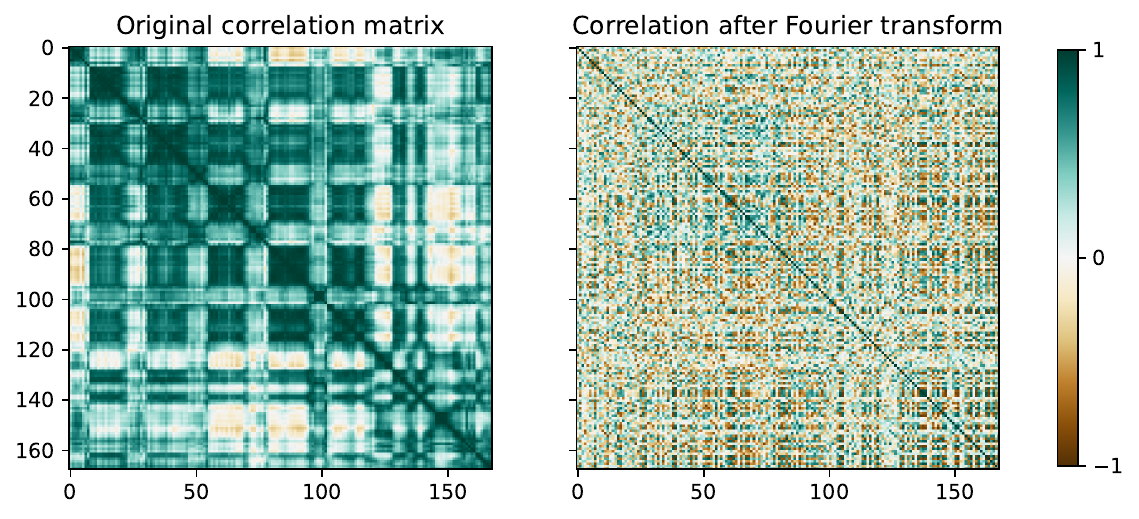}
    \caption{Time covariance matrix for the historical series shown in the top panel of Fig.~\ref{fig:sample-time-series} (left),
    and for their Fourier transform (right). Off-diagonal elements are larger, with clear structures reflecting periodicities for the time covariance matrix; they 
    are strongly suppressed for its Fourier transform.}
    \label{fig:correlation-matrix}
\end{figure}

The way this is done is rather involved, because one needs to generate new time series while keeping their
average, variance and periodicities (daily, weekly and seasonal), as well as realistic
correlations between load nodes. Daily and weekly periodicities are evident in the top panel of 
Fig.~\ref{fig:sample-time-series}, where we show four historical load curves for the fourth week of 2019 to 2022. They reflect themselves as periodic oscillations in the off-diagonal elements of the time  covariance matrix
\begin{equation}
    M(t_1,t_2) = \frac{1}{N} \sum_{i=1}^N L_i(t_1) L_i(t_2)
    - \frac{1}{N^2} \sum_{i=1}^N L_i(t_1) \sum_{j=1}^N L_j(t_2),
\end{equation}
where $L_i(t)$ indicates the load at bus $i$ as a function of the time $t$.
The matrix $M(t_1,t_2)$  corresponding to the data in the top panel of 
Fig.~\ref{fig:sample-time-series} (i.e. with $N=4$ different time series) is shown in
the left panel of Fig.~\ref{fig:correlation-matrix}. It exhibits clear structures in its off-diagonal elements, related to the intrinsic periodicities of the time series. 
Directly adding uncorrelated noisy fluctuations to the time series would inevitably erase these periodicities and generate unrealistic time series compared to real-world ones.
These periodicities can in principle be taken into account by  choosing appropriate noise sequences. However for the full, annual time series, this would force us to work 
with dense 8736$\times$8736 matrices, which is computationally time-consuming. One may however expect that, given the relatively few expected periodicities (daily, weekly and seasonal), the covariance matrix is much sparser in Fourier/frequency space. This is confirmed in the right panel of Fig.~\ref{fig:correlation-matrix}. Our method is therefore based on Fourier transforming load time series and adding noise on Fourier components in a way that preserves the datasets periodicities.
A visual validation of the method is given in the bottom panel of Fig.~\ref{fig:sample-time-series}, which shows synthetic time series generated by our algorithm. They look realistic, i.e. similar to the real ones shown in the top panel.
We sketch the method in the following. 

Our starting point are the annual time series of national historical load data from 2015 to 2023. 
As just discussed, it is not possible to treat separate time steps as independent variables, because there is a strong correlation between successive time steps, but also between successive days and weeks. 
We consider instead the Fourier transform ${\cal \tilde C}_\alpha(\omega)$ of the $\alpha^{\rm th}$ national load time series ${\cal C}_\alpha(t)$, which has a mean value $\mu(\omega) = \langle {\cal \tilde C}_\alpha(\omega) \rangle$ and a covariance matrix $\sigma(\omega,\omega') = \langle {\cal \tilde C}_\alpha(\omega) {\cal \tilde C}_\alpha(\omega') \rangle - \mu(\omega) \mu(\omega')$.
In Fourier space,  dominating frequencies are characterized by large covariance matrix elements, however the vast majority of frequencies can be treated as uncorrelated noise. This is clearly visible in Fig.~\ref{fig:correlation-matrix}. To boost numerical efficiency, we set to zero all off-diagonal entries of $\sigma$ that are below a certain threshold; this generates a sparse, tractable matrix. 

With the two quantities $\mu(\omega)$ and $\sigma(\omega, \omega')$, we define 
one Gaussian statistical ensemble per country. Different synthetic time series, each labeled by 
$i=1, \ldots n_s$,
are then constructed from each of these national ensembles. 
To make a long story short, one first generates new Fourier components 
$\tilde{\cal C}^{(i)}_\alpha(\omega) = \mu(\omega)  + \sum_{\omega'}
S(\omega,\omega') \, \xi(\omega')$, where 
$S(\omega,\omega')$ is defined by the matrix identity $[S^\top \, S](\omega,\omega') = \sigma(\omega,\omega') $, and
$\xi(\omega')$ is a vector with normally distributed random components with zero average and unit variance. 
It is straightforward to see that $\tilde{\cal C}^{(i)}_\alpha(\omega)$ has the same frequency covariance matrix as ${\cal \tilde C}_\alpha(\omega)$. Second, one Fourier-transforms back to the time domain, $\tilde{\cal C}^{(i)}_\alpha(\omega) \rightarrow {\cal C}^{(i)}_\alpha(t)$ 
to obtain a realistic time series with the same statistical properties as the original ones.
Such synthetic time series are shown in Fig.~\ref{fig:sample-time-series}.
The process of drawing synthetic series only requires multiplying very sparse matrices, combined with the fast Fourier transform algorithm. It is a numerically very efficient way of generating any number $n_s$ of synthetic time series.

Variance and correlations of load time series may vary from country to country. It is for instance 
reasonable to expect that larger countries have less variance, and correlation distributions that extend to lower values. In practice, the variance of a series is close to the sum of the variance of $\mu$ and the variance $\sigma$ over the 8736 time steps.
At the same time, the covariance between distinct series is approximately equal to the variance of $\mu$.
This means that the Pearson correlation between pairs of time series taken from the same distribution can be approximated as
\begin{equation}
    \langle \rho \rangle \approx \frac{\var(\mu)}{\var(\mu) + \var(\sigma)}
    \qquad
    \text{where}
    \qquad
    \var(\mu) = \frac{1}{T} \sum_{t=1}^T \mu_t^2 - \left( \frac{1}{T} \sum_{t=1}^T \mu_t \right)^2,
    \qquad
    \var(\sigma) = \frac{1}{T} \sum_{t=1}^T \sigma_{tt} - \frac{1}{T^2} \sum_{t, t'=1}^T \sigma_{tt'}.
\end{equation}
Therefore, the typical correlation $\langle \rho \rangle$ can be adjusted by scaling $\sigma$ for each individual country.
Choosing $\langle \rho \rangle = 0.8$, we obtain correlations that are distributed around this value as in Fig.~\ref{fig:correlations}.
These correlations are in very good agreement with confidential data provided to us by SwissGrid, the TSO for Switzerland.
The correlations between individual loads could arguably differ country by country, but in the absence of actual data, the choice using a central value of 0.8 in all countries is reasonable. 
As a consequence, the correlations among time series from distinct countries are smaller than in each country separately.
Our method can of course be adapted to generate statistically different synthetic datasets for different countries.

Besides the correlation, the other parameter that needs adjusting is the overall normalization of the synthetic time series. The overall power balance of the system must be respected, and given that our model features an expected production value for all the generators, the demand must match this expected production. Moreover, each bus of the model is assigned a proportion of its country's total load, as mentioned above, but there is still some freedom in sharing the total demand between the countries. This can be fixed by taking into account the flows of power through the borders. Since the ENTSO-E data contains precise hour-by-hour records, it would be easy to generate synthetic series using the same technique as for the loads. However, the import/export is actually quite strongly correlated with the nuclear production of each country, and it makes therefore more sense to use historical data matching with the nuclear series. 
Moreover, we choose to only constrain the annual total rather than the hour-by-hour border flow in order to leave enough room for the dispatch procedure.
Table~\ref{tab:border-flows} gives the average exported power (or imported when negative) for each country of the model and for each year between 2016 and 2020, as computed from the actual ENTSO-E data.
In summary, each country's total load is computed as the sum of the expected production of its dispatchable generators and of the actual production of its nuclear generators, minus the export value. The total load is then shared onto all the buses with a weight calculated from the population density, and each bus is assigned a distinct synthetic time series that distributes the load over the year.

\subsection*{Power generation dispatch}

To complete our datasets, we need to construct realistic time series for power generators, i.e. we need to develop a dispatch algorithm. As mentioned above, we  consider nuclear generators separately, since they produce at or close to their rated power outside of maintenance periods (see Fig.~\ref{fig:nuclear-series}). We therefore assign to them an historical production profile from the ENTSO-E transparency platform and take them as input to the dispatch OPF, together with the just discussed synthetic time series for loads.

All non-nuclear generators are then treated as dispatchable, and their output at every time step is determined using an OPF.~\cite{Chatzivasileiadis2018} We describe this algorithm in some detail.
The PanTaGruEl model comes with a generation cost for each individual generator that allows defining an economic OPF in which the objective function is the total production cost.
Once minimized under constraints on power ramping rates, rated power generation, thermal limits for flows on power lines and so forth, the optimization procedure determines the power produced by each generator, given the instantaneous power demand.
The economic OPF generically suffers from two main problems: 
\begin{itemize}
    \item
    Several flows are systematically fixed at the thermal limit of the corresponding power line, loading the latter at 100 \% of its capacity.
    This is a standard observation in constrained, non-convex optimization problems with large number of variables: the solution lies typically on the boundary of the feasibility region. 
    \item
    Performing an independent OPF at every time step results in an uncontrolled average power output for each individual generator. Obtaining the desired availability factor would require tuning the production costs in a subtle way, which is not feasible in practice.
\end{itemize}
To bypass these obstacles, we first formulate an OPF that includes collectively all time steps of a given year as variables of the optimization problem. In this way, we can enforce the availability factor as a constraint.
Second, we incorporate thermal limits on power lines not as rigid constraints, but include them into the objective function.
This approach is directly inspired by the actual way of working of TSOs, whose goal is to minimize the losses on the transmission network, but first and foremost to ensure the network's security by guaranteeing that all flows remain below the thermal limits, even in the case of failure of one or several elements of the grid, while minimizing the cost of out-of-market control actions such as redispatches.

Since our purpose is to dispatch power generation, it is sufficient to consider a DC power flow and work in the lossless line approximation.\cite{machowski2008power} In this case, the flow of power $\Phi_a$ on the line $a$ is related to the voltage angles $\theta_i$ at bus $i$ by a linear equation, given in vector notation by $\vec{\Phi} = \mathbf{B} \mathbf{M}^T \vec{\theta}$, where $\mathbf{M}$ is the incidence matrix of the network and $\mathbf{B}$ a diagonal matrix encoding the susceptance of all lines.
This flow is in turn related to the net power injections $P_i - L_i$  (generation minus load) at each bus through $\mathbf{M} \vec{\Phi} = \vec{P} - \vec{L}$.
The power flow equation $\vec{P} - \vec{L} = \mathbf{M} \mathbf{B} \mathbf{M}^T \vec{\theta}$ can therefore be inverted to get
\begin{equation}
    \vec{\Phi}(t) = \mathbf{B} (\mathbf{M} \mathbf{B})^\dagger
    \left[ \vec{P}(t) - \vec{L}(t) \right].
\end{equation}
where $(\mathbf{M} \mathbf{B})^\dagger$ is the pseudo-inverse of the matrix $\mathbf{M} \mathbf{B}$.
The OPF is then written as an optimization problem whose variables are the power outputs $P_i(t)$ of each generator $i$ at each time step $t$, and with an objective function quadratic in the $P_i(t)$ through the flow $\Phi_a$ on each line $a$,
\begin{equation}
    \min_{P_i(t)} \left\{ \sum_t \sum_a \left( \Phi_a^\text{thermal} \right)^{-1}
    \big[ \Phi_a(t) \big]^2
    + \sum_t \sum_i c_i(t) P_i(t)
    \right\} \, ,
    \label{eq:OPFobjective}
\end{equation}
subject to the constraints
\begin{equation}
    0 \leq P_i(t) \leq P_i^\text{rated},
    \qquad\qquad
    \sum_{i \in {\rm gens}} P_i(t) = \sum_{i \in {\rm loads}} L_i(t)
    \qquad\qquad
    \frac{1}{T} \sum_t P_i(t) = A_i P_i^\text{rated},
    \qquad\qquad
    \left| P_i(t) - P_i(t + 1) \right| \leq \Delta P_i^\text{ramp}.
    \label{eq:OPFconstraints}
\end{equation}
The constraints are respectively imposing that the power output of each generator is between zero and its rated capacity, that the total power produced at any time step $t$ is equal to the load $L_i(t)$ summed at all buses $i$, that the annual power output of each generator matches the availability determined in the model (availability factor $A_i$ multiplying the rated power), and finally that the ramping rate of the generator is respected. In our case, we only apply a maximum ramping rate of 200~MW per hour for coal-fired power plants.
The ramp constraints are taken to be periodic, in agreement with the approximate periodicity of the load profiles and variable generation costs.
Note that the second and third constraints are only compatible if the balance between the loads and the rated power weighted by the availability factor is respected.
This is ensured by the normalization of the loads discussed above.

\begin{figure}[ht]
    \centering
    (a) Verbano, 2016, week 44 \\
    \includegraphics[width=0.9\linewidth]{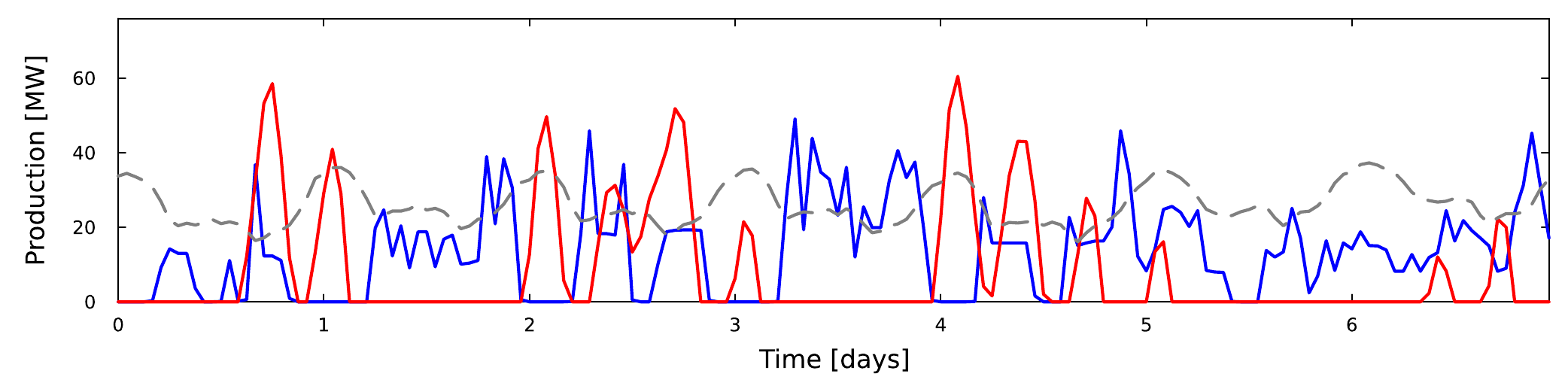} \\
    (b) Biasca, 2018, week 44 \\
    \includegraphics[width=0.9\linewidth]{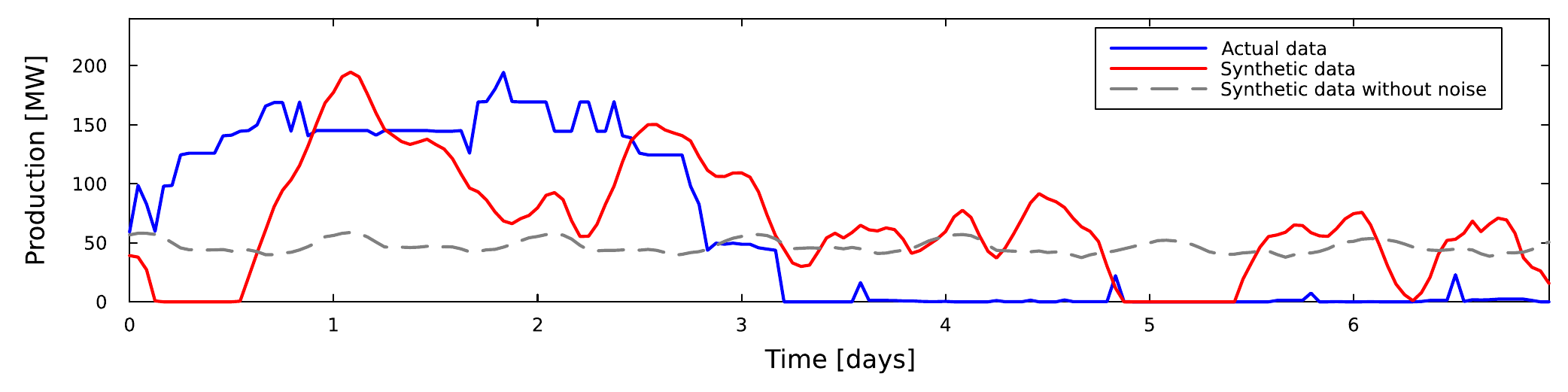}
    \caption{Comparison between synthetic time series and actual data
    for two Swiss hydroelectric plants, during a selected week in October.
    Different operating modes are clearly visible, and well captured by the synthetic data.
    The result of an optimal power flow without the noise component in the objective function
    is shown for comparison.}
    \label{fig:production-examples}
\end{figure}

The objective function, Eq.~\eqref{eq:OPFobjective}, contains two terms. The first term penalizes the lines that carry a large flow relative to their capacity, with a cost proportional to the line's capacity.
The second term represents varying generation costs. This term corresponds to the linear cost term in an economic OPF, except that the average value of $c_i(t)$ is irrelevant here: it only adds a constant to the objective function, because of the third constraint of a fixed annual production in Eq.~\eqref{eq:OPFconstraints}. The varying generation cost allows us to model the contingencies of a real production environment.
The time-dependent function $c_i(t)$ is taken to be a periodic noise defined as
\begin{equation}
    c_i(t) = \sqrt{\frac{2}{n}} \sum_{\nu = 1}^n \hat{A}_{i, \nu} \cos(2\pi \nu t + \hat{\theta}_{i, \nu})
    \label{eq:noise}
\end{equation}
where
\begin{itemize}
    \item we sum over frequencies $\nu$ that are annual, $\nu = i / (24 \times 364)$ with $i = 1, \ldots 10$, 
    weekly, $\nu = i / (24 \times 7)$ with $i = 1, \ldots 6$, and
    daily, $\nu = i / 24$ with $i = 1, \ldots 3$,
    \item $n=19$ is the total number of frequencies $\nu$,
    \item the amplitudes $\hat{A}_{i,\nu}$ are random variables obeying independent normal distributions with zero mean and unit variance,
    \item the phases $\hat{\theta}_{i, \nu}$ are random variables distributed uniformly in the interval $[0, 2\pi)$.
\end{itemize}
The effect of this superposition of harmonics with random amplitude and phase on the outcome of the OPF is illustrated in Fig.~\ref{fig:production-examples} for two hydro-storage generators. In the absence of noise, power generation is a relatively smooth function with weak fluctuations, because it only has to minimize flows on power lines. With noise, on the contrary, power generation exhibits fast, large-amplitude variations that are characteristic of hydro-storage plants. 

\begin{figure}[ht]
    \centering
    \includegraphics[width=0.49\linewidth]{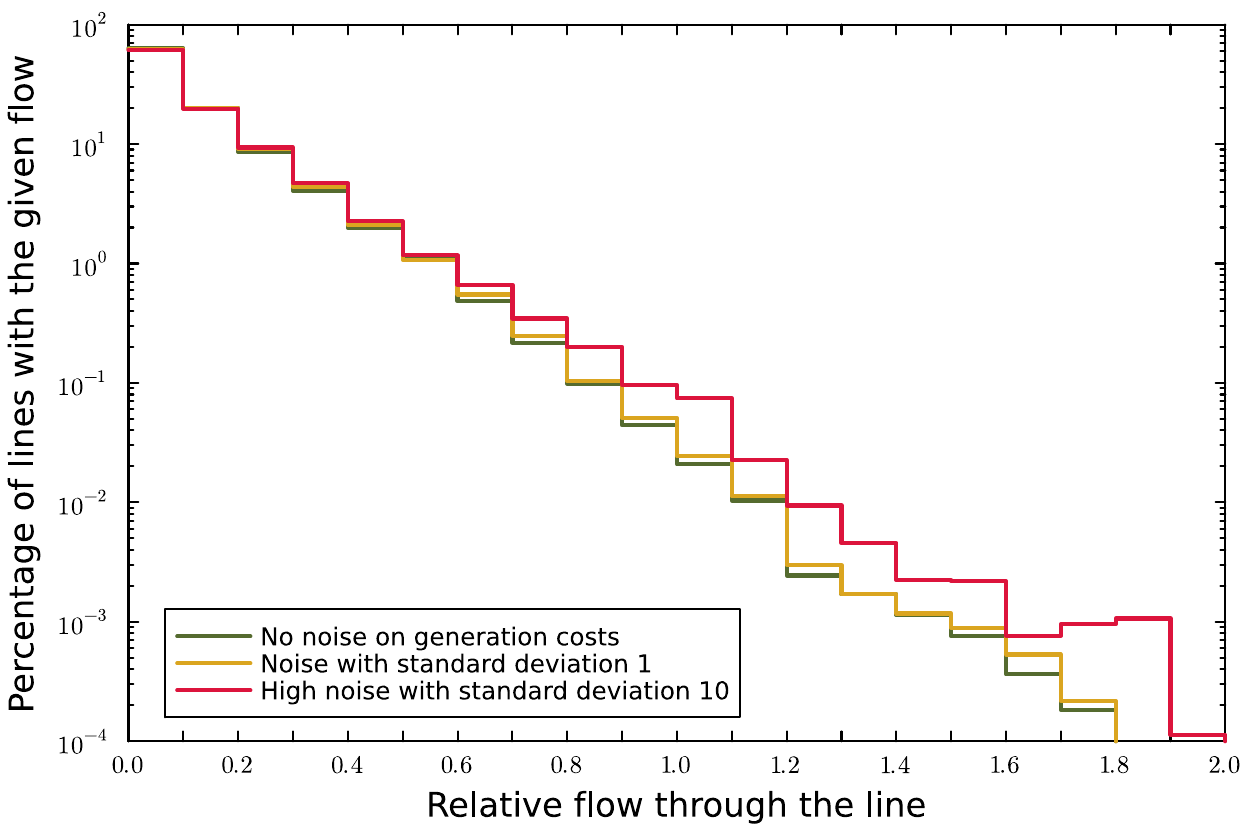}
    \hfill
    \includegraphics[width=0.49\linewidth]{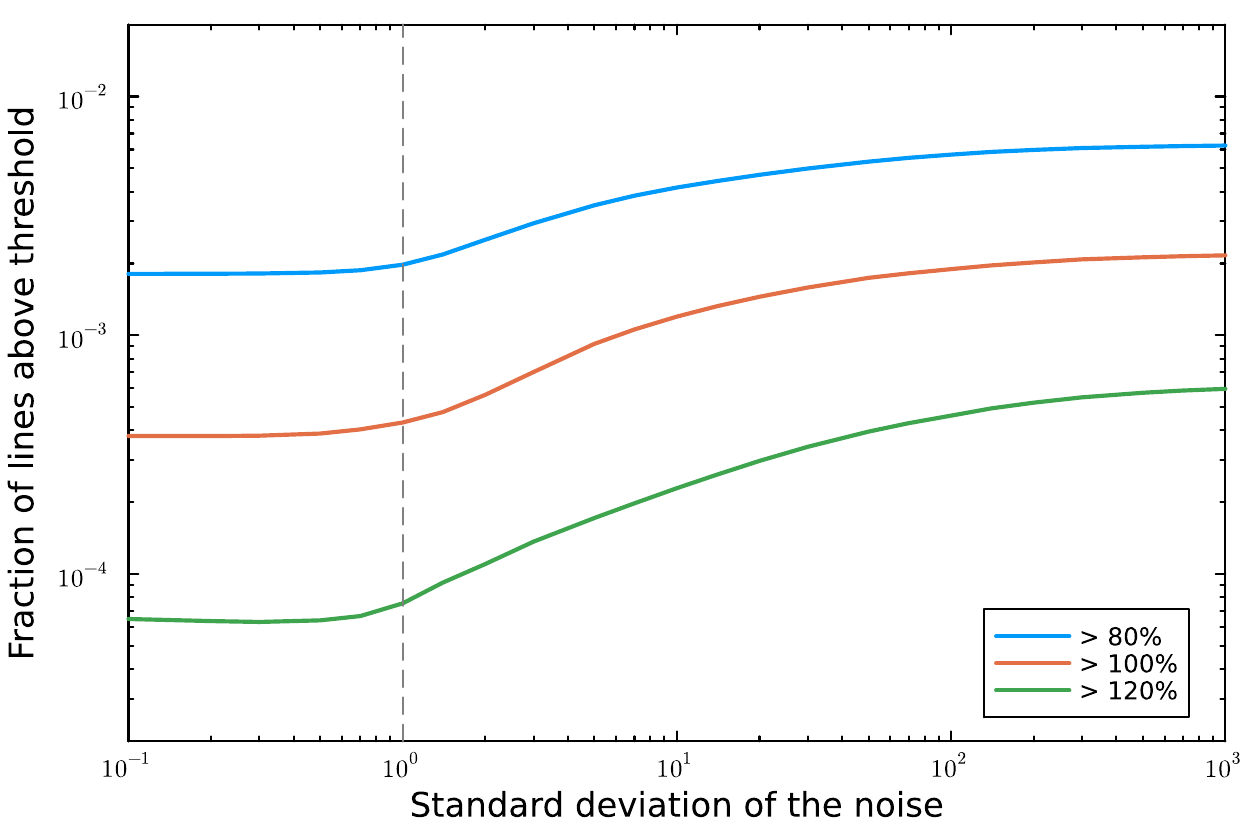}
    \caption{Behavior of the line flows as a function of the noise.
    The left panel shows how the tail of the distribution of line flows changes
    with the standard deviation of the noise:
    there is no significant difference between the unit standard deviation
    used to generate the data and a situation without noise, whereas the number of overloaded
    lines increases drastically if the noise is multiplied by a factor of 10.
    The right panel displays the fraction of lines above 80, 100, and 120~\%
    of their thermal limit, showing the transition between a regime dominated by the line cost
    (standard deviation of the noise below 1, vertical line)
    and a regime dominated by the noise (standard deviation of the noise above 100).}
    \label{fig:line-rates}
\end{figure}

The variance $\langle c_i^2(t) \rangle$ of the noise is the only free parameter in the optimization problem of Eqs.~\eqref{eq:OPFobjective} and \eqref{eq:OPFconstraints}.
The normalization of Eq.~\ref{eq:noise} is chosen so that $\langle c_i^2(t) \rangle = 1$. A larger variance
generates too many, too frequent overloaded lines, whereas smaller values result in smooth, constant production profiles that do not reflect market-induced dispatches, as seen in Fig.~\ref{fig:line-rates}.
Qualitative validation of the approach and of this choice for the noise variance
is given in Fig.~\ref{fig:production-examples}, which compares ENTSO-E to synthetic data for two Swiss hydroelectric power plant with different regimes. Synthetic data exhibit variations similar to real data, both in amplitude and frequency.

The optimization problem formulated in this way is convex and always admits a solution: a feasible point is provided by the constant generation case $P_i(t) = A_i P_i^\text{rated}$, and the global optimum can be reached numerically by applying an interior point method such as Newton-Raphson. 
We choose to work with the Gurobi solver,\cite{gurobi} because it is fast and provides free academic licenses. Computationally, the large size of the optimization problem leads to a memory issue: with 815 generators whose output must be determined over 8736 time steps, one needs to optimize over 7.12 million variables.
To address this issue we partition each year into smaller optimization problems.
First, we perform a coarse-grained optimization to distribute the production into 52 weeks. It takes as input loads and variable generation cost that are averaged over one-week intervals, and outputs a weekly average production for all generators. These output values define new availability factors for each generator and for all 52 weeks, that are then used in 52 independent fine-grained optimization problems with 168 time steps each, finally determining the hour-by-hour production profiles. To avoid obtaining generators that are completely turned off or produce at their rated power for an entire week, we restrict the power outputs of the coarse-grained optimization to 90~\% of the original feasible interval, replacing the first constraint in Eq.~\eqref{eq:OPFconstraints} with $0.1 \, A_i P_i^\text{rated} \leq P_i(t) \leq (0.9 + 0.1 \, A_i) P_i^\text{rated}$.
For ramp constraints, the consistency across successive weeks is guaranteed by imposing for the first and last time steps of the fine-grained optimization a constraint that is determined by the result of the coarse-grained problem. 
We stress that these partitioning-reconnecting procedures are motivated solely by the need to reduce the computing memory size required by the optimization problem. We have checked using different partitions that the result of the optimization problem is qualitatively independent of the partitioning, as long as the partitions remain large enough.

\section*{Data Records}

The dataset is available at Zenodo (\href{https://zenodo.org/records/13378476}{https://zenodo.org/records/13378476}).\cite{zenodo}
It contains CSV files with values representing time series for loads, generators, and lines, as well as a JSON file describing the European network in PowerModels format\cite{PowerModels},
as summarized in Table~\ref{tab:files}.
Each CSV file is a table with 8736 rows, one for each hourly time step of a 364-day year.
The number of columns depends on the type of table: there are 4097 columns in load files, 
815 for generators, and 8375 for lines.
Each column is described by a header corresponding to the element identifier in the network file.
All values are given in per-unit, both in the model file and in the tables, i.e.~they are multiples of a base unit taken to be 100 MW.

There are 20 tables of each type, labeled with a reference year (2016 to 2020). This amounts to a total of 20 years of synthetic data. The reference year determines the nuclear profiles used, as well as the import/export balance by country. The load profiles as well as the noise on the generation cost are distinct for all labels. When using load, generator, and lines profiles together, it is important to use the same label: for instance, the files \textit{loads\_2020\_1.csv}, \textit{gens\_2020\_1.csv}, and \textit{lines\_2020\_1.csv} represent a same year of the dataset, whereas \textit{gens\_2020\_2.csv} shares the same nuclear profiles, but it is based on a dispatch with distinct loads.

Note that the line information is redundant: it can be obtained from a DC power flow computation given the network and the injections (loads and generators). We provided it nevertheless for the user's convenience.

\section*{Technical Validation}

\begin{figure}[ht]
    \centering
    \includegraphics[width=0.49\linewidth]{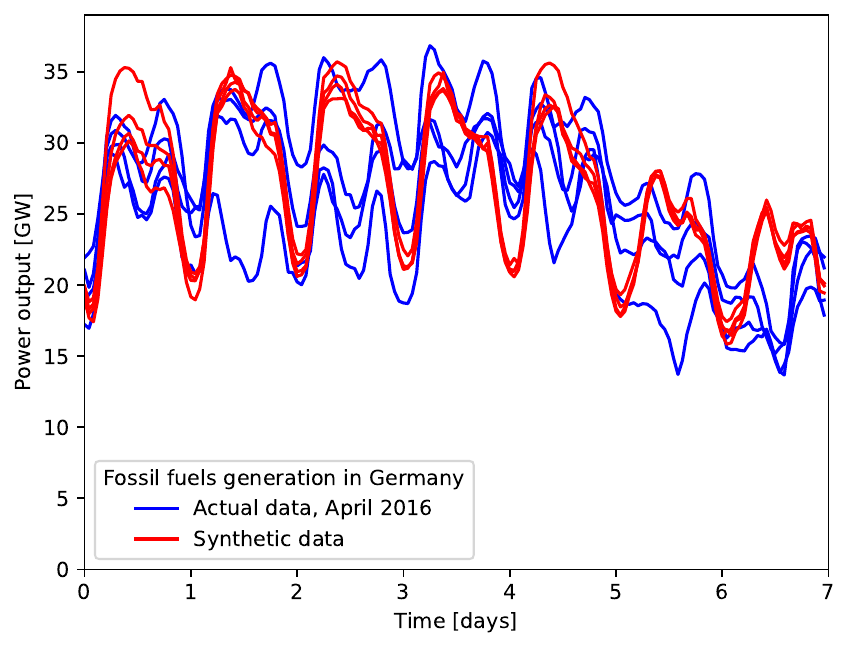}
    \hfill
    \includegraphics[width=0.49\linewidth]{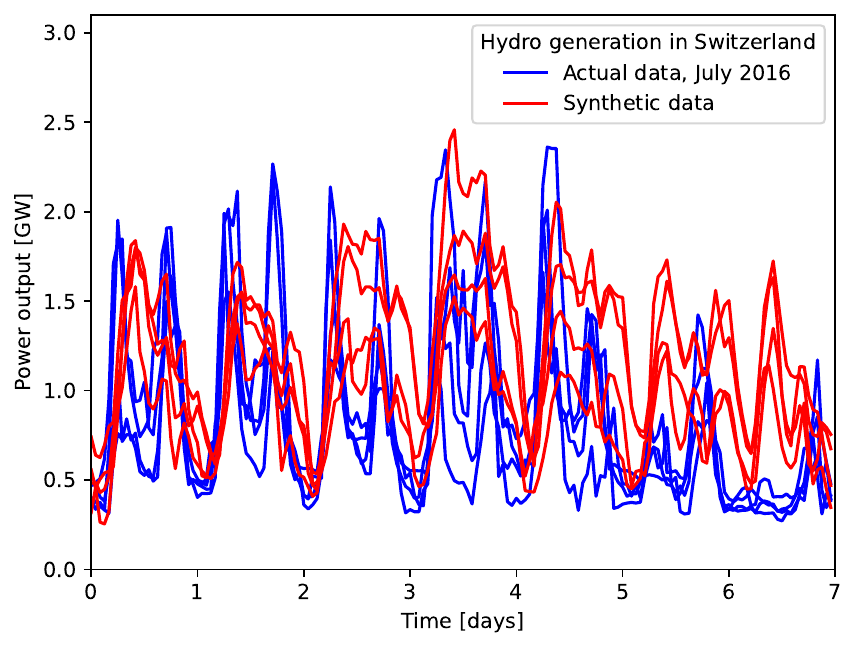}
    \caption{Comparison of our synthetic time series with actual production data,
    in two distinct situations: the left panel shows the aggregated production of fossil fuels
    in Germany during a spring month (coal, gas and oil, excluding nuclear); the right panel the aggregated hydroelectric production in Switzerland during a summer month. 4 weeks are superimposed in each case, giving a visual estimate of the variance both in the real and in the synthetic data.}
    \label{fig:production-aggregated}
\end{figure}

The validity of our dataset has be established in several ways:
\begin{itemize}
    \item The synthetic load profiles share all the characteristic features of real data.
    The typical annual, weekly and daily modulations are well-reproduced by the statistical model,  but also the amplitude and the frequency of the perturbations. This is clearly visible in Fig.~\ref{fig:sample-time-series}, and it is confirmed by a detailed spectral analysis of the time series.
    At the level of individual loads, each profile is drawn from the same distribution so that all the characteristics of the fluctuations are preserved.
    When aggregated country by country, the average total load matches the real-world data gathered from the ENTSO-E transparency platform.

    \item A crucial element for machine learning applications is the correlation between different time series. Realistic load profiles should be positively correlated among themselves because they result from the same physical situations, but they should never be perfectly correlated.
    We use as a metric the distribution of Pearson correlation coefficients between any pairs of load profiles, and compare our synthetic data with actual data obtained from Swissgrid, the Swiss TSO.
    As shown in Fig.~\ref{fig:correlations}, we obtain an excellent agreement.
    
    \item For production profiles, a direct comparison with actual ENTSO-E data has been performed on selected individual generators for which data is available. As illustrated in Fig.~\ref{fig:production-examples}, the apparently random behavior on short time scale is well reproduced.
    We observe also that different operating modes existing in the real world are present in the synthetic data: the generators are sometimes closely following the average load, but they can also operate in on/off mode.
    On longer time scales, the average production of the generators matches the real-life power stations: this is guaranteed by the availability constraint in our dispatching algorithm, and it has been verified in the resulting data.

    \item 
    The validation of the production profiles is also performed at the aggregated level.
    Grouping the power production by country and type, we check that our synthetic dataset agrees with aggregated data available from ENTSO-E.
    This is shown in Fig.~\ref{fig:production-aggregated} for two distinct production types: fossil fuels in Germany (excluding nuclear), and hydroelectric power in Switzerland. The two cases exhibit quite different behavior, as expected, but the qualitative and quantitative agreement between our dataset and the real-world measurements is clearly visible in both.
    This confirms that the dispatch is coherent in space.
\end{itemize}

\section*{Usage Notes}

All data in the repository is provided in human-readable formats, namely comma-separated values (CSV) for the time series and JavaScript Object Notation (JSON) for the network file.
It is ready to use in most frameworks (Python, Julia, Matlab, \ldots).
The format of the network file follows the PowerModels standard,\cite{PowerModels} which is itself derived from MatPower\cite{MatPower} and compatible with PandaPower.\cite{PandaPower}
It can therefore be used with the most popular tools in the field.
The identifier of the loads, generators, and lines match the column names of the corresponding CSV files.

For machine learning applications, the time series can be used without a reference to the network file, simply using all or a selection of columns of the CSV files, depending on the needs.
The dataset repository\cite{zenodo} contains a detailed description on how to select series from a particular country, or how to aggregate hourly time steps into days or weeks.
Note also that all of the yearly time series provided in the repository are periodic, which means that it is always possible to define a coherent time window modulo the length of the series.

\section*{Code availability}

The code used to generate the dataset as well as the figures and numbers quoted in this paper is made freely available in a GitHub repository (\href{https://github.com/GeeeHesso/PowerData}{https://github.com/GeeeHesso/PowerData}).
It consists in two packages and several documentation notebooks.
The first package, written in Python, provides functions to handle the data and to generate synthetic series based on historical data.
The second package, written in Julia, is used to perform the optimal power flow.
The documentation in the form of Jupyter notebooks contains numerous examples on how to use both packages. The entire workflow described in this manuscript is also provided, starting from raw ENTSO-E data files and ending with the synthetic dataset given in the repository.
This means for instance that arbitrarily many additional years of data can be generated, but also that the same methods can be applied to different power networks, or using other data sources.

\newpage

\bibliography{bibliography}

\newpage

\section*{Acknowledgements}

This work was supported by the Cyber-Defence Campus of armasuisse and by an internal research grant of the Engineering and Architecture domain of HES-SO. We thank Laurent Pagnier for discussion on the PanTaGruEl model of the synchronous grid of continental Europe and SwissGrid for access to sets of historical data.

\section*{Author contributions statement}

M.G. and G.D. collected the input data and constructed the dataset generation algorithm. P.J. provided advice on the methodology. 
M.G. and P.J. participated in writing the manuscript. All authors discussed the methods and the datasets. 

\section*{Competing interests}

The authors declare no competing interests.

\section*{Figures \& Tables}

\begin{table}[ht]
    \centering
    \footnotesize
    {\setlength{\tabcolsep}{3pt}
    \begin{tabular}{|c|r@{\hspace{1.5mm}}r@{\hspace{1.5mm}}r|r@{\hspace{1.5mm}}r@{\hspace{1.5mm}}r|r@{\hspace{1.5mm}}r@{\hspace{1.5mm}}r|r@{\hspace{1.5mm}}r@{\hspace{1.5mm}}r|r@{\hspace{1.5mm}}r@{\hspace{1.5mm}}r|r@{\hspace{1.5mm}}r@{\hspace{1.5mm}}r|r@{\hspace{1.5mm}}r@{\hspace{1.5mm}}r|rrr|}
        \hline
        & \multicolumn{3}{c|}{\multirow{2}{*}{\textbf{Nuclear}}}
        & \multicolumn{3}{c|}{\multirow{2}{*}{\textbf{Coal}}}
        & \multicolumn{3}{c|}{\multirow{2}{*}{\textbf{Gas and oil}}}
        & \multicolumn{9}{c|}{\textbf{Hydro}}
        & \multicolumn{3}{c|}{\multirow{2}{*}{\textbf{Other}}}
        & \multicolumn{3}{c|}{\multirow{2}{*}{\textbf{Total}}} \\
        & \multicolumn{3}{c|}{}
        & \multicolumn{3}{c|}{}
        & \multicolumn{3}{c|}{}
        & \multicolumn{3}{c|}{\textbf{storage}}
        & \multicolumn{3}{c|}{\textbf{river}}
        & \multicolumn{3}{c|}{\textbf{unspecified}} 
        & \multicolumn{3}{c|}{}
        & \multicolumn{3}{c|}{} \\
        & $N$ & $P_\text{rated}$ & \% av
        & $N$ & $P_\text{rated}$ & \% av
        & $N$ & $P_\text{rated}$ & \% av
        & $N$ & $P_\text{rated}$ & \% av
        & $N$ & $P_\text{rated}$ & \% av
        & $N$ & $P_\text{rated}$ & \% av
        & $N$ & $P_\text{rated}$ & \% av
        & $N$ & $P_\text{rated}$ & \% av \\
        \hline
        \textbf{AL} & && & && & 1 & 0.1 & 18.8 & 4 & 1.6 & 13.5 & 1 & 0.5 & 41.8 & && & && & \textbf{6} & \textbf{2.2} & \textbf{20.2} \\
		\textbf{AT} & && & 4 & 1.6 & 27.9 & 11 & 5.6 & 20.8 & 15 & 4.4 & 17.2 & 37 & 5.8 & 57.6 & && & 1 & 0.1 & 24.1 & \textbf{68} & \textbf{17.6} & \textbf{32.8} \\
		\textbf{BA} & && & 4 & 1.5 & 44.3 & && & && & 9 & 1.6 & 41.8 & && & && & \textbf{13} & \textbf{3.1} & \textbf{43.0} \\
		\textbf{BE} & 2 & 5.9 & 79.4 & 1 & 0.6 & 44.3 & 11 & 3.4 & 36.9 & 1 & 1.2 & 17.4 & && & && & 1 & 0.2 & 27.8 & \textbf{16} & \textbf{11.2} & \textbf{57.7} \\
		\textbf{BG} & 1 & 2.0 & 89.8 & 2 & 1.9 & 55.1 & 4 & 3.2 & 18.8 & 2 & 0.5 & 19.3 & 4 & 0.5 & 41.8 & 1 & 0.1 & 16.1 & && & \textbf{14} & \textbf{8.3} & \textbf{45.7} \\
		\textbf{CH} & 4 & 3.3 & 66.6 & && & && & 30 & 8.2 & 13.4 & 2 & 0.2 & 36.2 & && & && & \textbf{36} & \textbf{11.7} & \textbf{28.9} \\
		\textbf{CZ} & 2 & 3.8 & 64.2 & 13 & 6.5 & 47.7 & 1 & 0.4 & 37.4 & 2 & 1.1 & 14.4 & 2 & 0.5 & 28.1 & && & && & \textbf{20} & \textbf{12.2} & \textbf{48.6} \\
		\textbf{DE} & 9 & 12.1 & 84.6 & 46 & 49.5 & 49.4 & 26 & 12.5 & 9.9 & 17 & 8.2 & 10.4 & && & 1 & 0.1 & 21.2 & 2 & 0.8 & 17.5 & \textbf{101} & \textbf{83.2} & \textbf{44.4} \\
		\textbf{DK} & && & 3 & 1.1 & 22.4 & 7 & 3.2 & 15.4 & && & && & && & 1 & 0.1 & 26.0 & \textbf{11} & \textbf{4.4} & \textbf{17.5} \\
		\textbf{ES} & 2 & 3.2 & 84.2 & 3 & 1.9 & 42.3 & 3 & 2.5 & 17.8 & && & 14 & 4.8 & 88.4 & 29 & 5.7 & 17.0 & 10 & 7.1 & 24.1 & \textbf{61} & \textbf{25.3} & \textbf{43.3} \\
		\textbf{FR} & 19 & 63.1 & 69.0 & 1 & 0.8 & 30.8 & 22 & 16.3 & 37.9 & 5 & 3.1 & 21.1 & 64 & 12.0 & 46.6 & 4 & 0.2 & 32.3 & 1 & 0.6 & 15.8 & \textbf{116} & \textbf{96.1} & \textbf{58.7} \\
		\textbf{GR} & && & 7 & 5.2 & 44.2 & 14 & 7.0 & 29.8 & 13 & 2.8 & 13.5 & && & 2 & 0.3 & 21.9 & && & \textbf{36} & \textbf{15.3} & \textbf{31.5} \\
		\textbf{HR} & && & 1 & 0.3 & 44.3 & 3 & 0.9 & 18.8 & 5 & 1.3 & 13.5 & 2 & 0.2 & 41.8 & 2 & 0.2 & 21.9 & 1 & 0.3 & 26.6 & \textbf{14} & \textbf{3.1} & \textbf{21.0} \\
		\textbf{HU} & 1 & 1.9 & 91.4 & 4 & 0.6 & 59.7 & 14 & 5.1 & 13.9 & && & && & && & && & \textbf{19} & \textbf{7.6} & \textbf{36.9} \\
		\textbf{IT} & && & 7 & 11.7 & 14.4 & 47 & 49.7 & 16.0 & 13 & 2.9 & 9.4 & 9 & 7.3 & 34.0 & 1 & 0.1 & 21.0 & 2 & 0.3 & 86.0 & \textbf{79} & \textbf{71.9} & \textbf{17.6} \\
		\textbf{LU} & && & && & 1 & 0.4 & 41.3 & 1 & 1.1 & 18.0 & && & && & && & \textbf{2} & \textbf{1.5} & \textbf{24.1} \\
		\textbf{ME} & && & && & && & 1 & 0.3 & 13.6 & 1 & 0.3 & 41.8 & && & 1 & 0.2 & 26.6 & \textbf{3} & \textbf{0.9} & \textbf{26.9} \\
		\textbf{MK} & && & 2 & 0.7 & 33.5 & 2 & 0.4 & 19.7 & && & 4 & 0.4 & 41.8 & && & && & \textbf{8} & \textbf{1.6} & \textbf{31.9} \\
		\textbf{NL} & 1 & 0.5 & 87.8 & 7 & 5.2 & 77.9 & 17 & 9.3 & 20.3 & && & && & && & && & \textbf{25} & \textbf{15.0} & \textbf{42.7} \\
		\textbf{PL} & && & 28 & 23.2 & 54.5 & 7 & 1.7 & 43.9 & 4 & 1.5 & 4.8 & 1 & 0.2 & 45.2 & && & 3 & 0.2 & 22.9 & \textbf{43} & \textbf{26.7} & \textbf{50.7} \\
		\textbf{PT} & && & 2 & 1.8 & 76.1 & 5 & 3.9 & 28.4 & 17 & 4.4 & 20.0 & 11 & 2.2 & 38.0 & && & 4 & 0.3 & 30.6 & \textbf{39} & \textbf{12.5} & \textbf{33.9} \\
		\textbf{RO} & 1 & 1.4 & 99.0 & 11 & 5.1 & 31.4 & 9 & 3.2 & 22.5 & && & 23 & 5.1 & 48.8 & && & 1 & 0.4 & 21.4 & \textbf{45} & \textbf{15.2} & \textbf{41.3} \\
		\textbf{RS} & && & 8 & 5.2 & 44.3 & 2 & 0.3 & 18.8 & 2 & 1.1 & 13.5 & 2 & 1.1 & 41.8 & && & && & \textbf{14} & \textbf{7.7} & \textbf{38.6} \\
		\textbf{SI} & 1 & 0.7 & 88.8 & 2 & 0.3 & 55.5 & 1 & 0.3 & 8.5 & && & 6 & 0.5 & 47.2 & && & && & \textbf{10} & \textbf{1.7} & \textbf{58.2} \\
		\textbf{SK} & 2 & 1.9 & 86.6 & 3 & 1.1 & 44.2 & 2 & 0.6 & 11.9 & 2 & 0.8 & 3.7 & 1 & 0.7 & 38.1 & 3 & 0.2 & 20.0 & && & \textbf{13} & \textbf{5.5} & \textbf{47.5} \\
		\textbf{TR} & && & && & 3 & 1.4 & 18.8 & && & && & && & && & \textbf{3} & \textbf{1.4} & \textbf{18.8} \\
        \hline
		$\sum$ & \textbf{45} & \textbf{99.9} & \textbf{73.6} & \textbf{159} & \textbf{125.8} & \textbf{46.6} & \textbf{213} & \textbf{131.3} & \textbf{21.0} & \textbf{134} & \textbf{44.5} & \textbf{14.0} & \textbf{193} & \textbf{43.9} & \textbf{49.5} & \textbf{43} & \textbf{6.8} & \textbf{18.0} & \textbf{28} & \textbf{10.5} & \textbf{24.9} & \textbf{815} & \textbf{462.8} & \textbf{41.4} \\
        \hline
    \end{tabular}}
    \caption{Production by generator type and country for the PanTaGruEl model of the synchronous grid of continental Europe, giving the number
    $N$ of independent generators, the total rated power $P_\text{rated}$ of these $N$ generators in GW, and their average availability factor in percent. 
    For instance, France (country code "FR") has 19 nuclear generators
    with a total capacity of 63~GW and an availability of 69 \%. }
    \label{tab:model-gens}
\end{table}

\begin{table}[ht]
    \centering
    \small
    \begin{tabular}{|c|c|c|c|c|c||c|c|c|c|c|}
        \hline
        & \multicolumn{5}{c||}{Average exported power in MW}
        & \multicolumn{5}{c|}{National load scaling factor} \\
        & \textbf{2016} & \textbf{2017} & \textbf{2018} & \textbf{2019} & \textbf{2020}
        & \textbf{2016} & \textbf{2017} & \textbf{2018} & \textbf{2019} & \textbf{2020} \\
		\hline
		\textbf{AL} & $+$5 & $-$331 & $+$113 & $-$274 & $-$255
		& 0.10 & 0.18 & 0.08 & 0.17 &  0.16\\
		\textbf{AT} & $-$994 & $-$891 & $-$1182 & $-$504 & $-$380
		& 0.96 & 0.94 & 0.99 & 0.89 & 0.87 \\
		\textbf{BA} & $+$426 & $+$206 & $+$523 & $+$541 & $+$515
		& 0.73 & 0.90 & 0.65 & 0.64 & 0.66 \\
		\textbf{BE} & $-$726 & $-$736 & $-$2001 & $-$413 & $-$526
		& 0.74 & 0.73 & 0.71 & 0.71 & 0.62 \\
		\textbf{BG} & $+$190 & $+$375 & $+$661 & $+$449 & $+$222
		& 0.82 & 0.76 & 0.71 & 0.76 & 0.82 \\
		\textbf{CH} & $-$513 & $-$731 & $-$114 & $+$378 & $+$482
		& 0.56 & 0.58 & 0.58 & 0.52 & 0.50 \\
		\textbf{CZ} & $+$1251 & $+$1510 & $+$1555 & $+$1450 & $+$1110
		& 0.64 & 0.66 & 0.68 & 0.70 & 0.74 \\
		\textbf{DE} & $+$5663 & $+$6004 & $+$5875 & $+$3982 & $+$2456
		& 0.55 & 0.53 & 0.55 & 0.58 & 0.61 \\
		\textbf{DK} & $-$282 & +131 & $-$165 & $-$394 & +473
		& 0.27 & 0.17 & 0.24 & 0.30 & 0.08 \\
		\textbf{ES} & $-$1454 & $-$1714 & $-$1666 & $-$699 & $-$415
		& 0.49 & 0.49 & 0.50 & 0.46 & 0.45 \\
		\textbf{FR} & $+$3016 & $+$3160 & $+$5365 & $+$5266 & $+$4010
		& 1.03 & 1.02 & 1.01 & 0.98 & 0.91 \\
		\textbf{GR} & $-$838 & $-$341 & $-$387 & $-$835 & $-$794
		& 0.99 & 0.90 & 0.91 & 0.99 & 0.98 \\
		\textbf{HR} & $-$718 & $-$838 & $-$695 & $-$700 & $-$573
		& 0.68 & 0.74 & 0.67 & 0.67 & 0.61 \\
		\textbf{HU} & $-$964 & $-$961 & $-$1071 & $-$1015 & $-$1168
		& 0.76 & 0.76 & 0.77 & 0.77 & 0.80 \\
		\textbf{IT} & $-$4133 & $-$4164 & $-$4889 & $-$4493 & $-$3792
		& 0.51 & 0.51 & 0.54 & 0.52 & 0.50 \\
		\textbf{LU} & 0 & 0 & $-$482 & $-$471 & $-$431
		& 0.67 & 0.67 & 1.59 & 1.57 & 1.49 \\
		\textbf{ME} & $-$31 & $-$124 & $+$24 & $-$124 & $-$71
		& 0.72 & 0.98 & 0.57 & 0.97 & 0.83 \\
		\textbf{MK} & $-$137 & $-$212 & $-$217 & $-$206 & $-$255
		& 0.15 & 0.17 & 0.17 & 0.16 & 0.18 \\
		\textbf{NL} & $-$991 & $-$650 & $-$1188 & $-$601 & $+$330
		& 0.60 & 0.56 & 0.61 & 0.56 & 0.49 \\
		\textbf{PL} & $+$235 & $+$272 & $-$63 & $-$469 & $-$665
		& 0.69 & 0.69 & 0.71 & 0.73 & 0.74 \\
		\textbf{PT} & $+$573 & $+$303 & $+$298 & $-$389 & $-$170
		& 0.64 & 0.69 & 0.69 & 0.81 & 0.78 \\
		\textbf{RO} & $+$664 & $+$518 & $+$330 & $+$49 & $-$8
		& 0.81 & 0.84 & 0.86 & 0.90 & 0.94 \\
		\textbf{RS} & $+$167 & $-$154 & $-$72 & $-$56 & $+$78
		& 0.66 & 0.73 & 0.71 & 0.71 & 0.68 \\
		\textbf{SI} & $+$128 & $+$52 & $+$48 & $+$47 & $+$231
		& 0.58 & 0.63 & 0.64 & 0.64 & 0.52 \\
		\textbf{SK} & $-$537 & $-$685 & $-$600 & $-$522 & $-$402
		& 0.97 & 1.02 & 0.99 & 0.97 & 0.90 \\
		\hline
    \end{tabular}
    \caption{Average power imported ($-$) or exported ($+$) by all countries of the European model (left), and scaling factors applied to the national load so that the balance Eq.~\eqref{eq:sumrule} is respected.
    The scaling factor can be significantly below 1 for countries that have few generators 
    directly connected to the ultra-high-voltage transmission grid.}
    \label{tab:border-flows}
\end{table}

\begin{table}[ht]
    \centering
    \footnotesize
    \begin{tabular}{|c|c|c|c|}
        \hline
        \multicolumn{4}{|c|}{\large \textbf{Time series}} \\
        \hline
        \textbf{Archive files} & \textbf{Individual files} & \makecell{\textbf{Border flows \&} \\ \textbf{nuclear profiles}} & \textbf{Description} \\
        \hline
        \texttt{loads\_2016.zip}
        & \makecell{\texttt{loads\_2016\_1.csv}~~
        \texttt{loads\_2016\_2.csv} \\
        \texttt{loads\_2016\_3.csv}~~
        \texttt{loads\_2016\_4.csv}} & 2016 &
        \makecell[l]{\textbf{Columns}: 4097 loads (bus label as heading) \\
        \textbf{Rows}: 8736 time steps (364 days $\times$ 24 hours)} \\
        \cline{1-3}
        \texttt{loads\_2017.zip}
        & \makecell{\texttt{loads\_2017\_1.csv}~~
        \texttt{loads\_2017\_2.csv} \\
        \texttt{loads\_2017\_3.csv}~~
        \texttt{loads\_2017\_4.csv}} & 2017 &
        \makecell[l]{\textbf{Values}: Active power consumption ($\times$100 MW)} \\
        \cline{1-3}
        \texttt{loads\_2018.zip}
        & \makecell{\texttt{loads\_2018\_1.csv}~~
        \texttt{loads\_2018\_2.csv} \\
        \texttt{loads\_2018\_3.csv}~~
        \texttt{loads\_2018\_4.csv}} & 2018 & 
        \makecell[l]{All consumption time series are distinct, each \\
        generated independently by the algorithm.} \\
        \cline{1-3}
        \texttt{loads\_2019.zip}
        & \makecell{\texttt{loads\_2019\_1.csv}~~
        \texttt{loads\_2019\_2.csv} \\
        \texttt{loads\_2019\_3.csv}~~
        \texttt{loads\_2019\_4.csv}} & 2019 & \\
        \cline{1-3}
        \texttt{loads\_2020.zip}
        & \makecell{\texttt{loads\_2020\_1.csv}~~
        \texttt{loads\_2020\_2.csv} \\
        \texttt{loads\_2020\_3.csv}~~
        \texttt{loads\_2020\_4.csv}} & 2020 & \\
        \hline
        \texttt{gens\_2016.zip}
        & \makecell{\texttt{gens\_2016\_1.csv}~~
        \texttt{gens\_2016\_2.csv} \\
        \texttt{gens\_2016\_3.csv}~~
        \texttt{gens\_2016\_4.csv}} & 2016 & 
        \makecell[l]{\textbf{Columns}: 815 generators (labels as heading) \\
        \textbf{Rows}: 8736 time steps (364 days $\times$ 24 hours)} \\
        \cline{1-3}
        \texttt{gens\_2017.zip}
        & \makecell{\texttt{gens\_2017\_1.csv}~~
        \texttt{gens\_2017\_2.csv} \\
        \texttt{gens\_2017\_3.csv}~~
        \texttt{gens\_2017\_4.csv}} & 2017 & 
        \makecell[l]{\textbf{Values}: Generator's active power output ($\times$100 MW)} \\
        \cline{1-3}
        \texttt{gens\_2018.zip}
        & \makecell{\texttt{gens\_2018\_1.csv}~~
        \texttt{gens\_2018\_2.csv} \\
        \texttt{gens\_2018\_3.csv}~~
        \texttt{gens\_2018\_4.csv}} & 2018 &
        \makecell[l]{All production time series are distinct, each resulting  \\
        from an OPF with the corresponding load profiles} \\
        \cline{1-3}
        \texttt{gens\_2019.zip}
        & \makecell{\texttt{gens\_2019\_1.csv}~~
        \texttt{gens\_2019\_2.csv} \\
        \texttt{gens\_2019\_3.csv}~~
        \texttt{gens\_2019\_4.csv}} & 2019 &
        \makecell[l]{and different noise on generation costs. \\ ~} \\
        \cline{1-3}
        \texttt{gens\_2020.zip}
        & \makecell{\texttt{gens\_2020\_1.csv}~~
        \texttt{gens\_2020\_2.csv} \\
        \texttt{gens\_2020\_3.csv}~~
        \texttt{gens\_2020\_4.csv}} & 2020 & \\
        \hline
        \texttt{lines\_2016.zip}
        & \makecell{\texttt{lines\_2016\_1.csv}~~
        \texttt{lines\_2016\_2.csv} \\
        \texttt{lines\_2016\_3.csv}~~
        \texttt{lines\_2016\_4.csv}} & 2016 & 
        \makecell[l]{\textbf{Columns}: 8375 lines (labels as heading) \\
        \textbf{Rows}: 8736 time steps (364 days $\times$ 24 hours)} \\
        \cline{1-3}
        \texttt{lines\_2017.zip}
        & \makecell{\texttt{lines\_2017\_1.csv}~~
        \texttt{lines\_2017\_2.csv} \\
        \texttt{lines\_2017\_3.csv}~~
        \texttt{lines\_2017\_4.csv}} & 2017 & 
        \makecell[l]{\textbf{Values}: Active power flowing through the line} \\
        \cline{1-3}
        \texttt{lines\_2018.zip}
        & \makecell{\texttt{lines\_2018\_1.csv}~~
        \texttt{lines\_2018\_2.csv} \\
        \texttt{lines\_2018\_3.csv}~~
        \texttt{lines\_2018\_4.csv}} & 2018 &
        \makecell[l]{All time series are distinct, each resulting from a \\
        power flow computation with the corresponding.} \\
        \cline{1-3}
        \texttt{lines\_2019.zip}
        & \makecell{\texttt{lines\_2019\_1.csv}~~
        \texttt{lines\_2019\_2.csv} \\
        \texttt{lines\_2019\_3.csv}~~
        \texttt{lines\_2019\_4.csv}} & 2019 & 
        \makecell[l]{injections (generators and loads). \\ ~} \\
        \cline{1-3}
        \texttt{lines\_2020.zip}
        & \makecell{\texttt{lines\_2020\_1.csv}~~
        \texttt{lines\_2020\_2.csv} \\
        \texttt{lines\_2020\_3.csv}~~
        \texttt{lines\_2020\_4.csv}} & 2020 & \\
        \hline
        \hline
        \multicolumn{4}{|c|}{\large \textbf{Other files}} \\
        \hline
        \multicolumn{2}{|c|}{\textbf{File name}} &
        \multicolumn{2}{c|}{\textbf{Description}} \\
        \hline
        \multicolumn{2}{|l|}{\texttt{europe\_network.json}} &
        \multicolumn{2}{l|}{\makecell[l]{Network file containing a list of buses, lines, generators and loads,  in \\
        PowerModels format,\cite{PowerModels} serialized as JSON text file.}} \\
        \hline
        \multicolumn{2}{|l|}{\texttt{loads\_by\_country.csv}} &
        \multicolumn{2}{l|}{\makecell[l]{List of loads for each country in the model. \\
        \textbf{Columns}: two-letter country code. \textbf{Values}: bus labels.}} \\
        \hline
        \multicolumn{2}{|l|}{\texttt{gens\_by\_country.csv}} &
        \multicolumn{2}{l|}{\makecell[l]{List of generators for each country in the model. \\
        \textbf{Columns}: two-letter country code. \textbf{Values}: generator labels.}} \\
        \hline
    \end{tabular}
    \caption{List of files in the Zenodo data repository.\cite{zenodo}
    There are 20 independent time series for production and consumption,
    amounting to 20 years of synthetic data. In each case, we provide the power injections
    (generators and loads) as well as the flows on the lines.}
    \label{tab:files}
\end{table}

\end{document}